\documentclass[superscriptaddress,reprint,aps,prl]{revtex4-2}

\usepackage{graphicx}
\usepackage{amsmath}
\usepackage{amssymb}
\usepackage{dcolumn}
\usepackage{bm}
\usepackage{multirow}

\newcommand\Tstrut{\rule{0pt}{1.3em}}

\begin{document}


\title{Extracting and Storing Energy From a Quasi-Vacuum on a Quantum Computer}

\author{Songbo Xie}

\affiliation{Department of Chemistry, Department of Electrical and Computer
Engineering, and Purdue Quantum Science and Engineering Institute,
Purdue University, West Lafayette, 47907, Indiana, USA}
\affiliation{Department of Electrical and Computer Engineering, North Carolina State University, Raleigh, North Carolina 27606, USA }

\author{Manas Sajjan}
\affiliation{Department of Chemistry, Department of Electrical and Computer
Engineering, and Purdue Quantum Science and Engineering Institute,
Purdue University, West Lafayette, 47907, Indiana, USA}
\affiliation{Department of Electrical and Computer Engineering, North Carolina State University, Raleigh, North Carolina 27606, USA }

\author{Sabre Kais}
\email{skais@ncsu.edu}
\affiliation{Department of Chemistry, Department of Electrical and Computer
Engineering, and Purdue Quantum Science and Engineering Institute,
Purdue University, West Lafayette, 47907, Indiana, USA}
\affiliation{Department of Electrical and Computer Engineering, North Carolina State University, Raleigh, North Carolina 27606, USA }

\date{\today}

\begin{abstract}
We explore recent advancements in the understanding and manipulation of vacuum energy in quantum physics, with a focus on the quantum energy teleportation (QET) protocol. Traditional QET protocols extract energy from what we refer to as a ``quasi-vacuum'' state, but the extracted quantum energy is dissipated into classical devices, limiting its practical utility. To address this limitation, we propose an enhanced QET protocol that incorporates an additional qubit, enabling the stored energy to be stored within a quantum register for future use. We experimentally validated this enhanced protocol using IBM superconducting quantum computers, demonstrating its feasibility and potential for future applications in quantum energy manipulation.
\end{abstract}

\maketitle
{\it Introduction.---}In physics, it is generally understood that a ``vacuum'' is not completely empty \cite{milonni2019introduction}. Fluctuations of a quantum vacuum can generate virtual particles, leading to phenomena such as spontaneous emission \cite{dirac1927quantum,weisskopf1935probleme}, vacuum polarization \cite{brown1996lowest,levine1997measurement}, the Casimir effect \cite{casimir1948influence,lamoreaux1997demonstration}, and the Unruh effect \cite{fulling1973nonuniqueness,davies1975scalar,unruh1976notes}. Furthermore, the vacuum state of a relativistic free field is known to exhibit entanglement \cite{summers1985vacuum}. This vacuum entanglement can even be harvested, allowing a pair of particles that have never directly interacted to become entangled long before they are causally related \cite{reznik2005violating,sabin2010dynamics,sabin2012extracting}. 

More importantly, a vacuum has what is known as zero-point energy, the lowest possible energy that a quantum system can have. Studying how to extract and store the vacuum energy for future use is both fundamentally fascinating and practically valuable. Various methods have been proposed to extract energy from the vacuum, including electrical rectification of the vacuum field \cite{mead1996system,valone2009proposed}, mechanical extraction using Casimir cavities \cite{forward1984extracting,widom1998casimir,pinto2005method}, and pumping atoms with Casmir cavities \cite{haisch2008quantum}. However, none of these approaches have been experimentally realized. Concerns and debates persist regarding their feasibility \cite{cole1993extracting,abbott1997quantum,little2006null,moddel2019extraction}.

Fortunately, a new breakthrough for extracting ``zero-point vacuum energy'' has emerged. In 2008, Masahiro Hotta proposed a novel protocol using quantum information science to extract energy from a ``vacuum'' \cite{hotta2008protocol}. This protocol, known as {\it quantum energy teleportation} (QET), has been followed by numerous theoretical studies involving various quantum systems \cite{hotta2009quantumchain,hotta2009quantumtrapped,nambu2010quantum,hotta2010quantum,yusa2011quantum,trevison2015quantum}. Very recently, QET has been realized experimentally using nuclear magnetic resonance (NMR) systems \cite{rodriguez2023experimental} and IBM superconducting quantum computers \cite{ikeda2023demonstration}.

We must clarify that the energy extraction enabled by QET is not from a true vacuum, but from what we will argue is a ``quasi-vacuum,'' which shares similar properties with a true vacuum. Additionally, the energy extraction process is not free: it requires spending energy from a distant location, making it appear as if energy is being teleported from one place to another. Furthermore, communication between the two distant locations is necessary, preventing the protocol from being superluminal. These characteristics collectively define the protocol as ``quantum energy teleportation'' (QET).

However, the original QET protocol has a limitation: the energy extracted from the vacuum is lost into a classical device and cannot be used for future purposes. Therefore, developing methods to store and harness the extracted energy is essential for enhancing the QET protocol's effectiveness and utility.

\begin{figure}[t]
    \centering
    \includegraphics[width=\linewidth]{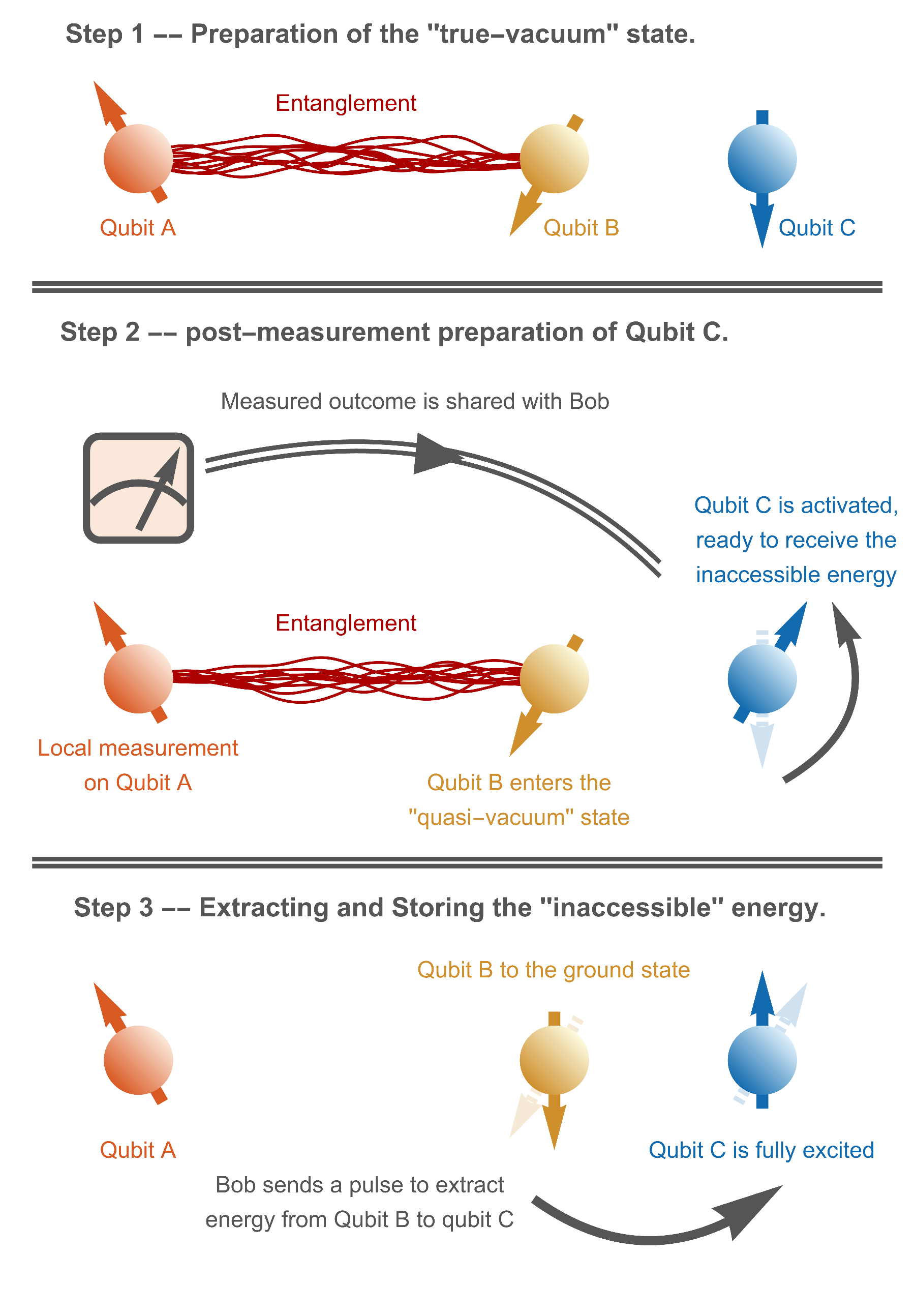}
    \caption{The process of the enhanced QET protocol. Step 1: Alice and Bob prepare the three-qubit ground state based on the Hamiltonian $H_A + H_V + H_B + H_C$. Step 2: Alice performs a local measurement on qubit A and communicates the outcome to Bob. Bob then utilizes this information to activate qubit C, making it ready to receive the energy within qubit B. Step 3: Using Alice's information again, Bob sends a pulse to extract energy from qubit B, which is then stored in qubit C for future use.}
    \label{fig:process}
\end{figure}

In this work, we address this limitation of QET by introducing an enhanced QET protocol that includes an additional qubit. By performing a post-measurement state preparation for the additional qubit, we demonstrate that one can successfully extract energy from a quasi-vacuum state and store it in the additionally prepared qubit for future use. A quick overview of the new protocol's process is given by Fig.~\ref{fig:process}. To validate the feasibility of our enhanced protocol, we experimentally implement it using IBM superconducting quantum computers.

\begin{figure}[t]
    \centering
    \includegraphics[width=1\linewidth]{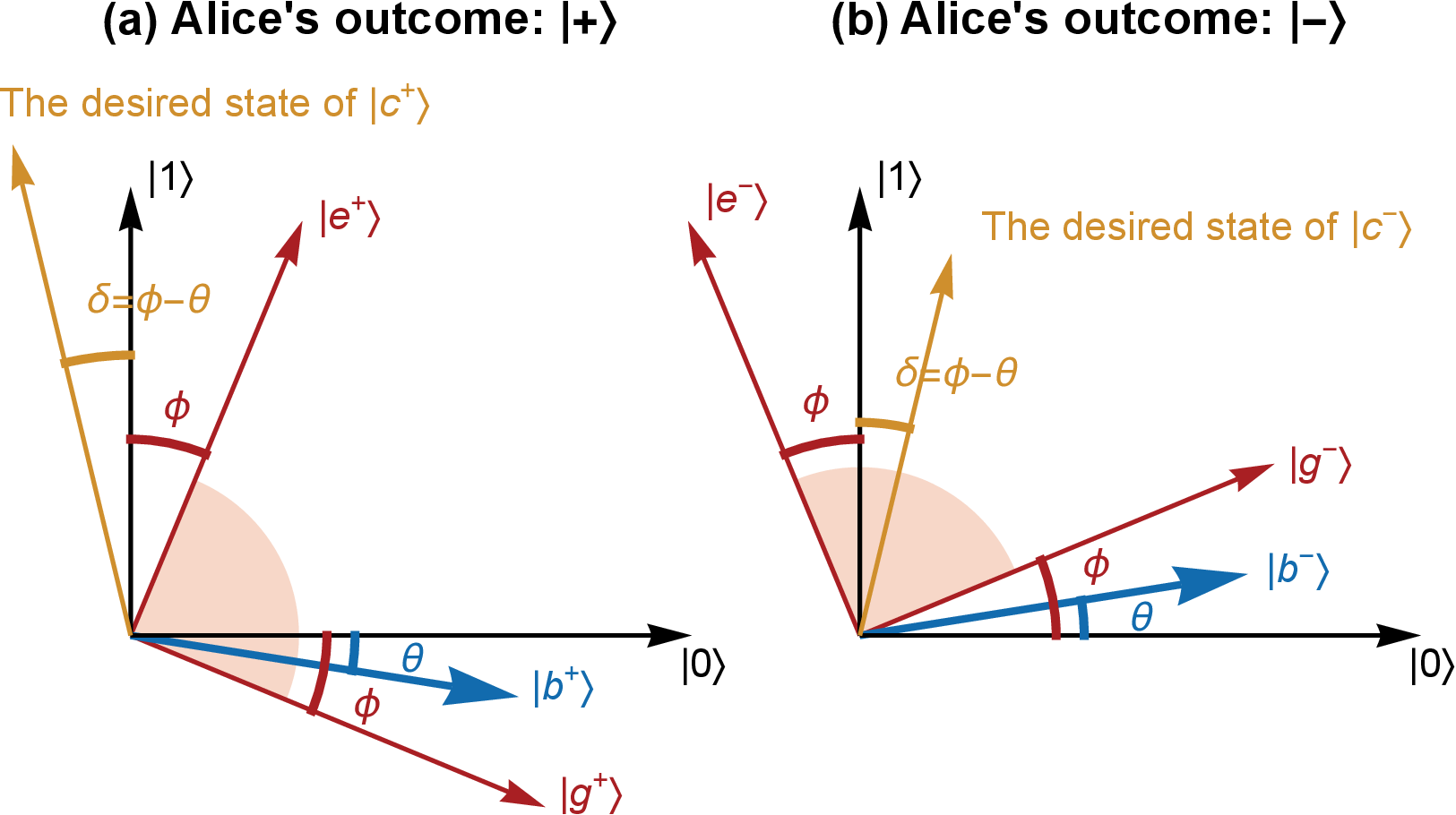}
    \caption{(a) If Alice's measurement outcome is ``$+$'', the effective Hamiltonian for qubit $B$ becomes $H^{(+)}_\text{eff}$, with the dressed eigenstates $|e^+\rangle$ and $|g^+\rangle$, as defined in Eq.~\eqref{dressed}. In this case, the desired state for qubit $C$ is $|c^+\rangle$. (b) When Alice's measurement outcome is ``$-$'', the effective Hamiltonian for qubit $B$ changes to $H^{(-)}_\text{eff}$, with the dressed eigenstates $|e^-\rangle$ and $|g^-\rangle$. The desired state for qubit $C$ in this case is $|c^-\rangle$.}
    \label{fig:coordinate}
\end{figure}

{\it Reinterpretation of the original QET protocol---}For the development of the enhanced protocol, we reinterpret the original QET protocol by introducing an effective Hamiltonian given by Eq.~\eqref{dressed}. This new interpretation is further illustrated in Fig.~\ref{fig:coordinate}. 

To start with, it should be emphasized that the energy extracted by the QET protocol is not obtained from a true vacuum state. In specific, true vacuum states discussed in this work are ground states of given Hamiltonians, since we are working within the context of quantum information. In this framework, these true vacuum states are Strong Passive, meaning that no general quantum operation $\mathcal{G}$, represented by positive operator-valued measurements (POVM), can extract energy from the system. This is expressed by the nonnegativity of $\Delta E=\text{Tr}[H(\mathcal{G}|g\rangle\langle g|)]-\langle g|H|g\rangle\geq0$, where $H$ is the Hamiltonian of the system and $|g\rangle$ is the ground state. 

Instead, QET considers a different scenario: a multipartite system and in a Strong Local Passive (SLP) state, denoted by $\rho$. In such cases, no general quantum operation $\mathcal{G}$ applied locally to a subsystem can extract energy from the entire system. This is expressed by the nonnegativity of $\Delta E=\text{Tr}[H(\mathcal{I}\otimes\mathcal{G})\rho]-\text{Tr}(H\rho)\geq0$. We will demonstrate that these SLP states behave as what we refer to as ``quasi-vacuum'' states.

To illustrate, we consider the minimal model of QET involving two qubits $A$ and $B$, shared by Alice and Bob. The Hamiltonian is given by: $H_{AB} = H_A + H_B + H_{V}$, where $H_i = -h_i\sigma_z^i + f_i\mathbb{I}^i$ representing individual Hamiltonian for qubit $i$ with $i\in\{A,B\}$. $H_{V} = 2\kappa\sigma_x^A \otimes \sigma_x^B + f_{V}\mathbb{I}^{AB}$ represents the interaction Hamiltonian between the two qubits. $h_A, h_B, \kappa$, are positive constants and  $f_A, f_B, f_{V}$ are constants chosen such that the ground state $|g\rangle$ of the total Hamiltonian $H_{AB}$ has vanishing expectation values for all three terms: $\langle g|H_A|g\rangle = \langle g|H_B|g\rangle = \langle g|H_{V}|g\rangle = 0$. The expressions of these constants are derived in the supplemental material.

The two qubits are initially prepared in the ground state $|g\rangle$ of the $H_{AB}$, given by:
\begin{equation}\label{g1}
    \begin{split}
        &|g\rangle=\cos(\theta)|00\rangle_{AB} \; - \;\sin(\theta)|11\rangle_{AB},
    \end{split}
\end{equation}
where $\tan(\theta)\equiv\sqrt{(x_A+x_B)^2+1}-(x_A+x_B)$, with $x_A\equiv h_A/2\kappa$ and $x_B\equiv h_B/2\kappa$. Neither Alice nor Bob can extract any energy from $|g\rangle$ since a ground state has the lowest possible energy. 

Instead, the QET protocol requires Alice to perform a generalized measurement on her local qubit using Kraus operators $M_i$ that commute with the interaction Hamiltonian, i.e., $[M_i,H_{V}] = 0$. This requirement ensures two things. First, although Alice's measurement injects energy into her local term $H_A$, altering the system's state from the ground state, the energy terms associated with Bob remain unaffected, keeping $\langle H_B \rangle=\langle H_{V} \rangle=0$. Second, after Alice's measurement, the system becomes an SLP state, meaning Bob's local operations still cannot extract any energy. From Bob's viewpoint, the following facts can be observed: Bob's local energy, $\langle H_V+H_B\rangle=0$, remains equal to the true-vacuum energy; the outgoing energy flow is locally blocked for Bob. For these reasons, we refer to the SLP state, obtained by Alice's local measurement, as a ``quasi-vacuum'' state for Bob.

For Bob to extract the blocked energy from the quasi-vacuum state, the QET protocol requires Alice to share her measurement outcome with Bob. Based on these outcomes, Bob performs corresponding local operations on his qubit, allowing for the extraction of the previously blocked energy.

Specifically, Alice measures whether qubit $A$ is in the state $|+\rangle$ or $|-\rangle$, where $|\pm\rangle\equiv(|0\rangle\pm|1\rangle)/\sqrt{2}$ are the eigenstates of the Pauli-X matrix $\sigma_x$. The projective operators $|+\rangle\langle+|$ and $|-\rangle\langle-|$ commute with the Hamiltonian $H_V$, satisfying the requirement stated above. Accordingly, we can express Eq.~\eqref{g1} in a more compact form:
\begin{equation}\label{g2}
    |g\rangle=\dfrac{1}{\sqrt{2}}(|+\rangle_A|b^+\rangle_B+| \;-\;\rangle_A|b^-\rangle_B),
\end{equation}
where $|b^\pm\rangle\equiv\cos(\theta)|0\rangle\mp\sin(\theta)|1\rangle$ are Bob's local states contingent on Alice's outcomes. Note that $|b^+\rangle$ and $|b^-\rangle$ are not orthogonal.

Alice's measurement leads to a two-qubit state represented by the density matrix:
\begin{equation}\label{quasi}
    \rho_\text{quasi}=(|+\rangle\langle+|_A|b^+\rangle\langle b^+|_B \;+\; |-\rangle\langle-|_A|b^-\rangle\langle b^-|_B)/2.
\end{equation}
It can be shown that this state is an SLP state for Bob, from which Bob's local operations cannot extract energy. We call the state Eq.~\eqref{quasi} an ``quasi-vacuum'' state for Bob, with reasons explained earlier.

\begin{figure*}[t]
    \centering
    \includegraphics[width=0.8\linewidth]{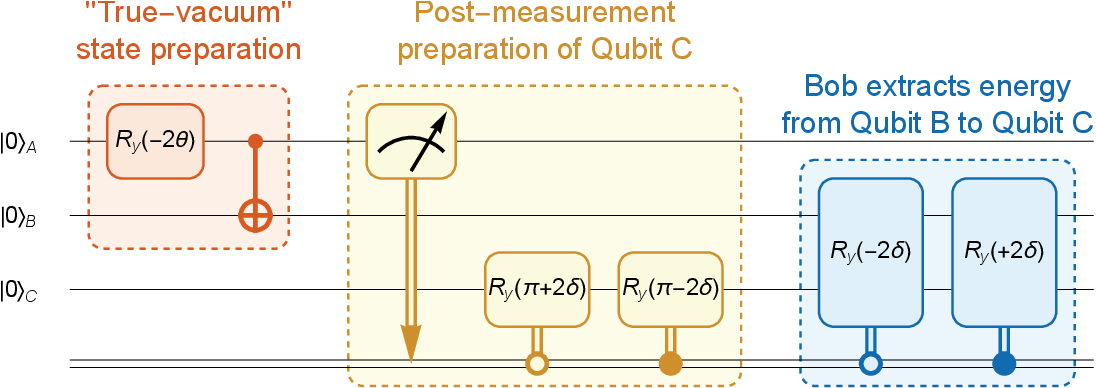}
    \caption{Quantum circuit design for the enhanced QET protocol. Step 1: To prepare the ``true-vacuum'' state of the Hamiltonian, a local rotation is applied to qubit $A$, followed by a CNOT gate between qubits $A$ and $B$. Step 2: Alice measures qubit $A$ and communicates the outcome to Bob. Bob then applies a conditional rotation to qubit $C$ to activate it for receiving energy from qubit $B$. Step 3: Based on Alice's outcome, Bob sends a pulse to qubits $B$ and $C$, extracting energy from qubit $B$ and storing it in qubit $C$.}
    \label{fig:circuit}
\end{figure*}

Generally, determining the expectation value of the interacting Hamiltonian, i.e. $\langle H_V\rangle$, requires knowledge of the states of both qubits $A$ and $B$. Here, since Alice's measurement commutes with $H_V$, the role of qubit $A$ in $H_V$ is fixed after her measurement. Specifically, the outcome of Alice's measurement (either $+$ or $-$) modifies Bob's Hamiltonian $(H_B+H_V)$ into an effective form: 
\begin{equation}\label{dressed}
    \begin{split}
        {H}_\text{eff}^{(\pm)}\equiv& H_V^{(\pm)}+H_B\\
        =&\pm2\kappa\sigma_x^B-h_B\sigma_z^B+(f_V+f_B)\mathbb{I}^B.
    \end{split}
\end{equation}
Given Bob's effective Hamiltonian, Bob's ``dressed eigenstates'' are expressed as:
\begin{equation}
    \begin{split}
        |e^\pm\rangle=\pm\sin(\phi)|0\rangle+\cos(\phi)|1\rangle,\\
        |g^\pm\rangle=+\cos(\phi)|0\rangle\mp\sin(\phi)|1\rangle,
    \end{split}
\end{equation}
where $\tan(\phi)\equiv\sqrt{x_B^2+1}-x_B$. The upper (lower) sign corresponds Alice's outcome $+$($-$). The eigenvalues of the effective Hamiltonian \eqref{dressed} are $h_C\equiv\sqrt{h_B^2+4\kappa^2}$ for $|e^\pm\rangle$ and $-h_C$ for $|g^\pm\rangle$. A detailed derivation can be found in the supplemental material.

In terms of these dressed states, Bob's local states $|b^\pm\rangle$ in Eq.~\eqref{g2} can be expressed as:
\begin{equation}
    |b^\pm\rangle\equiv\cos(\delta)|g^\pm\rangle\pm\sin(\delta)|e^\pm\rangle,
\end{equation}
with $\delta\equiv\phi-\theta$. Consequently, to extract energy from the ``quasi-vacuum'' state in Eq.~\eqref{quasi}, Bob must perform a conditional operation, $R_y(\mp2\delta)$, depending on Alice's outcome (either $+$ or $-$). These operations rotate his local states $|b^\pm\rangle$ to the corresponding dressed ground states $|g^\pm\rangle$. This procedure is illustrated by the coordinate systems in Fig.~\ref{fig:coordinate}.

Eq.~\eqref{quasi} has been described as a quasi-vacuum state for Bob, which prevents him from extracting energy through local operations. However, it is now evident that this energy-flow restriction can be removed if Bob knows Alice's measurement outcome and performs a conditional operation based on this information. This enables the extraction of energy that was previously inaccessible. This entire process is known as quantum energy teleportation (QET).

To practically perform such a unitary operation in experiments, Bob needs to apply an electromagnetic pulse to qubit $B$. According to the law of energy conservation, the energy extracted from qubit $B$ is transferred to this pulse. However, since the pulse is classical, the extracted energy is essentially lost and can be neglected, as the small portion of quantum energy is insignificant compared to the energy of a classical pulse. Therefore, even though it is fundamentally interesting to extract previously inaccessible energy from a quasi-vacuum state in the QET protocol, it is not practically useful if the extracted energy cannot be utilized. This raises a natural question: can we store this extracted energy in a quantum register for future use?

{\it Developing an enhanced QET protocol.---}Here, we provide a definitive answer to the above question. For the purpose of storing the quantum energy, Bob must prepare a third qubit $C$, which remains noninteracting with qubits $A$ and $B$, and has an independent Hamiltonian $H_C\equiv -h_C\sigma_z^C+h_C\mathbb{I}^C$. The term $h_C\mathbb{I}$ guarantees that the smallest eigenvalue of $H_C$ is 0. Since the state of qubit $B$ already depends on Alice's outcome, it is desirable for the state of qubit $C$ to also be influenced by Alice's outcome, allowing for energy transfer from qubit $B$ to qubit $C$ in each case. 

To achieve this, a common approach is to prepare the state of qubit $C$ before Alice's measurement. This requires Alice and Bob to prepare a three-qubit entangled state that extends beyond the form of Eq.~\eqref{g2}. The desired three-qubit state should be:
\begin{equation}\label{g3}
|\psi\rangle=\frac{1}{\sqrt{2}}(|+\rangle_A|b^+\rangle_B|c^+\rangle_C + |-\rangle_A|b^-\rangle_B|c^-\rangle_C),
\end{equation}
where $|c^\pm\rangle$ are the desired states for qubit $C$, contingent on Alice's outcome, and their specific forms will be determined later.

However, preparing Eq.~\eqref{g3} is challenging because $|b^+\rangle$ and $|b^-\rangle$ are not orthogonal, preventing Bob from using controlled-unitary gates to prepared $|c^\pm\rangle$ locally. Instead, to obtain Eq.~\eqref{g3}, Alice must prepare $|c^\pm\rangle$ on her side by performing controlled-unitary gates with qubit $A$ as the control, since $|+\rangle$ and $|-\rangle$ are orthogonal. She then uses the quantum state teleportation to transfer the states $|c^\pm\rangle$ to qubit $C$ in Bob's hand \cite{bennett1993teleporting}. However, quantum state teleportation requires Alice and Bob to share an additional entangled Bell state beforehand, which is very resource-expensive.

Instead of the approach described above, we propose that Bob performs a post-measurement state preparation for qubit $C$. Specifically, Bob initializes qubit $C$ in the state $|0\rangle$, such that the three qubits $A$, $B$, and $C$ are in the ground state $|\tilde{g}\rangle$ of the three-qubit Hamiltonian $H_{ABC}=H_{AB}+H_C$:
\begin{equation}
    |\tilde{g}\rangle=\dfrac{1}{\sqrt{2}}(|+\rangle_A|b^+\rangle_B \;+\; |-\rangle_A|b^-\rangle_B)\otimes|0\rangle_C.
\end{equation}
Bob then waits for Alice to perform the projective measurement $\{|+\rangle\langle+|,|-\rangle\langle-|\}$ as required by the original QET protocol and to send the outcome to Bob. Upon receiving Alice's message, Bob applies the conditional operation $R_y(\pi\pm2\delta)$ to transform the state of qubit $C$ to the desired states $|c^\pm\rangle\equiv\mp\sin(\delta)|0\rangle+\cos(\delta)|1\rangle$, depending on Alice's message $\pm$. This completes the preparation stage of our enhanced QET protocol by activating qubit $C$ for receiving energy in the next step. To be specific, if Alice's outcome is $+$, the state of Bob's qubits becomes $|b^+\rangle_B|c^+\rangle_C$. If Alice's outcome is $-$, the state of Bob's qubits becomes $|b^-\rangle_B|c^-\rangle_C$. See Fig.~\ref{fig:coordinate} for illustrations.

At this stage, the reduced density matrix of qubit $B$ remains identical to those in Eq.~\eqref{g1} and \eqref{g2}, indicating that energy extraction from qubit $B$ remains locally blocked for Bob. However, with Alice's message, Bob can perform conditional operations to extract the energy from qubit $B$. With the existence of the additional qubit $C$, we now show that the same operation can simultaneously store the extracted energy in qubit $C$. 

Suppose Alice's outcome is $+$, Bob sends an electromagnetic pulse to perform the $R_y(-2\delta)$ rotation on qubits $B$ and $C$. This pulse drives qubit $B$ to its dressed ground state $|g^+\rangle$ and drives qubit $C$ to its excited state $|1\rangle$. During this process, qubit $B$ loses an amount of energy $2\sin^2(\delta)h_C$, while qubit $C$ gains the same amount of energy. Conversely, if Alice's outcome is $-$, Bob sends an electromagnetic pulse to perform the $R_y(2\delta)$ rotation on both qubits, also resulting in the transfer of energy from qubit $B$ to qubit $C$ with the same amount of $2\sin^2(\delta)h_C$. See the supplemental material for detailed derivations.

According to the conservation law of energy, the classical pulse itself neither gains nor loses energy. Instead, the energy extracted from qubit $B$ is entirely transferred to qubit $C$. In summary, with Alice's message, Bob can always extract previously inaccessible energy from the ``quasi-vacuum'' state and store this energy to fully excited qubit $C$. This stored energy can then be used for future purposes.

{\it Experimental verification.---}To demonstrate the feasibility of our enhanced QET protocol, we implemented it on the IBM computer hardware ``ibm\_brisbane.'' There are three Hamiltonian parameters one can control: $h_A$, $h_B$, and $\kappa$. Due to homogeneity, we set $h_A=1$ without loss of generality. Numerical analyses indicate that to maximize the extracted and stored energy $2\sin^2(\delta)h_C$, one should set $\kappa=0.393$ and minimize the positive number $h_B$. Here, we set $h_B=0.01$, resulting in an nearly optimal extracted energy of the amount 0.295. With these chosen parameters, we have the following values: $f_A=0.789$, $f_B=0.008$, $f_V=0.483$, and $h_C=0.786$.

\begin{table}[t]
    \centering
    \begin{tabular}{c|c|ccc}
    \hline\hline
        \multicolumn{2}{c|}{Backends}& theory & Aer\_sim & ibm\_brisbane \Tstrut\\
        \hline
        \multirow{3}{*}{Step 1\ \rule{0pt}{1.8em}}& $\langle H_A\rangle$ & 0 & 0.001 & $-0.055\pm0.002$ \Tstrut\\
        & $\langle H_V+H_B\rangle$ & 0 & 0.001 & $-0.087\pm0.010$ \Tstrut\\
        & $\langle H_C\rangle$ & 0 & 0.0 & $-0.139\pm0.004$ \Tstrut\\
        \hline
        \multirow{3}{*}{Step 2\ \rule{0pt}{1.8em}} & $\langle H_A\rangle$ & 0.789 & 0.788 & $0.820\pm0.032$ \Tstrut\\
        & $\langle H_V+H_B\rangle$ & 0 & $-0.003$ & $-0.001\pm0.183$ \Tstrut\\
        & $\langle H_C\rangle$ & 1.277 & 1.271 & $1.314\pm0.022$ \Tstrut\\
        \hline
        \multirow{3}{*}{Step 3\ \rule{0pt}{1.8em}} & $\langle H_A\rangle$ & 0.789 & 0.793 & $0.780\pm0.017$ \Tstrut\\
        & $\langle H_V+H_B\rangle$ & $-0.295$ & $-0.295$ & $-0.397\pm0.137$ \Tstrut\\
        & $\langle H_C\rangle$ & 1.572 & $1.572$ & $1.670\pm0.078$ \Tstrut\\
        \hline
        \multicolumn{2}{c|}{Extracted $\Delta (E_V+E_B)$} & $-0.295$ & $-0.292$ & $-0.396\pm0.229$ \Tstrut\\
        \multicolumn{2}{c|}{Stored energy $\Delta E_C$} & 0.295 & 0.301 & $0.356\pm0.081$ \Tstrut\\
        \hline\hline
    \end{tabular}
    \caption{Experimental results from the quantum circuit executed on both the Aer Simulator and ibm\_brisbane are compared with theoretical values. The entire process is divided into three steps, with the energy of each qubit measured individually at each step. Finally, we calculate the energy extracted from qubit $B$ and the energy stored in qubit $C$ using our protocol. Further details, including the source of uncertainty in the measurements, can be found in the supplemental material.}
    \label{tab:result}
\end{table}

We divide the protocol into three steps, measuring the energy of all three qubits at each step. In step 1, Alice and Bob prepare the ``true-vacuum'' state of the Hamiltonian $H_{ABC}$, entangling qubits $A$ and $B$. In step 2, Alice performs the measurement $\{|+\rangle\langle+|,|-\rangle\langle-|\}$ on qubit $A$ and communicates the outcome to Bob. Based on this information, Bob rotates qubit $C$ into either the state $|c^+\rangle$ or $|c^-\rangle$. This activates qubit $C$ to receive energy from qubit $B$. Finally, in step 3, Bob sends one pulse to qubits $B$ and $C$. This pulse neither gains nor loses energy. Instead, it facilitates the extraction of energy from qubit $B$ and its transfer to qubit $C$. The quantum circuit for these three steps is shown in Fig.~\ref{fig:circuit}. A detailed decomposition of the circuit can be found in the supplemental material.

We run the quantum circuit on both the Aer Simulator and the ibm\_brisbane quantum hardware, comparing the results to theoretical values. The Aer Simulator, a classical simulator provided by the Qiskit package in Python, performs sampling for the given circuit with high accuracy. The ibm\_brisbane backend, on the other hand, is IBM's quantum computer with 127 superconducting qubits.

IBM quantum hardware, such as ibm\_brisbane, now supports dynamic circuits like the one shown in Fig.~\ref{fig:circuit}. Dynamic circuits allow for conditional operations based on classical measurement outcomes. However, in practice, we find that these dynamic circuits yield results with poor accuracy. This issue arises because dynamic circuits can take significantly longer time to execute than standard unitary circuits, often exceeding the coherence time of a qubit. To improve the accuracy of our experiments, we avoided using dynamic circuits and instead replaced them with equivalent circuits as discussed in \cite{ikeda2023demonstration}. See the supplemental material for more details.

The results are summarized in Table \ref{tab:result}. In step 1, the three qubits are initialized in their ground state, with all three Hamiltonian terms equal to 0. In step 2, Alice performs a measurement, injecting an energy of $0.789$ into $\langle H_A\rangle$, while Bob's energy---$\langle H_V+H_B\rangle$---remains unaffected at 0. Upon receiving Alice's information, Bob activates qubit $C$, injecting an energy of $1.277$ into $\langle H_C\rangle$. In step 3, Bob sends a signal that extracts an energy of $0.295$ from qubit $B$ and transfer it to qubit $C$, which fully excites qubit $C$. The results from both the Aer Simulator and the ibm\_brisbane hardware demonstrate the three steps with measured energies that closely match theoretical values, as shown in Table \ref{tab:result}. These well-aligned experimental results demonstrate the feasibility of our proposed protocol to extract and store quantum energy that was previously inaccessible within a ``quasi-vacuum'' state.

{\it Summary.---}While there is ongoing debate about whether energy can be extracted from a true-vacuum state, a recent breakthrough has introduced the quantum energy teleportation (QET) protocol, extracting energy from ``quasi-vacuum'' states---states that exhibit strong local passivity (SLP). However, a key limitation of the QET protocol is that the extracted energy is lost to a classical device, making the protocol less useful when that energy cannot be effectively harnessed. To overcome this limitation, we propose an enhanced QET protocol by introducing an additional qubit. By performing post-measurement state preparation on this qubit, we activate it to receive and store the quantum energy extracted via the original QET protocol, making the stored energy available for future use. We demonstrate the feasibility of our protocol by running quantum circuits on IBM quantum hardware and a classical simulator, comparing the results with theoretical predictions. The high accuracy of these results confirms the practicality of our enhanced QET protocol. In the future, it would be valuable to implement the protocol on other quantum platforms, such as NMR systems, to further validate its feasibility.

{\it Acknowledgments.---}We would like to thank Professor Joseph Eberly for valuable discussions. We acknowledge funding from the Office of Science through the Quantum Science Center (QSC), a National Quantum Information Science Research Center, and the U.S. Department of Energy (DOE) (Office of Basic Energy Sciences), under Award No. DE-SC0019215. 

\bibliography{extract}
\end{document}